# MEASURING SPIRAL ARM TORQUES:
# RESULTS FOR M100

Oleg Y. Gnedin, Jeremy Goodman, and Zsolt Frei

Princeton University Observatory, Peyton Hall, Princeton, NJ 08544

Electronic mail: ognedin@astro.princeton.edu, jeremy@astro.princeton.edu, frei@astro.princeton.edu

# ABSTRACT

Spiral arms, if they are massive, exert gravitational torques that transport angular momentum radially within galactic disks. These torques depend not on the pattern speed or permanence of the arms but only on the nonaxisymmetric mass distribution. Hence the torques can be measured from photometry. We demonstrate this using *gri* CCD data for M100 (NGC 4321). Since we find consistency among the three bands, we believe that dust and young stars in the arms do not seriously bias our results. If the present epoch is representative, the timescale for redistribution of angular momentum in M100 is $5 - 10$ Gyr, the main uncertainty being the mass-to-light ratio of the disk.

## 1. Introduction

It has long been recognized that spiral structure may be associated with secular dynamical changes in galactic disks. The anti-spiral theorem of Lynden-Bell & Ostriker (1967) states that strictly steady state spiral modes of linear amplitude cannot exist without dissipation, except where they are superpositions of degenerate nonspiral modes. *Nonlinear* selfconsistent models of spirals have been sought [Contopoulos & Grosbøl (1988), Contopoulos (1990) and references therein]. Lynden-Bell & Kalnajs (1972) have shown, however, that massive spiral arms transmit gravitational torques. If the arms trail, then the torques transport angular momentum outward through the disk. Linear or nonlinear, dissipative or conservative, a rigorously steady spiral pattern therefore requires sources and sinks of angular momentum. But is this a problem only of principle; in other words, how significant are spiral torques compared to the ages and angular momenta of their galactic hosts?

On the basis of extensive photographic photometry, Schweizer (1976) argued that red (or rather orange) light traces the old stellar disk in external spirals. If this is true then spiral arms represent an enhancement in the surface density of mass. Since Schweizer's influential study, many workers have Fourier analyzed photometry of spirals in red or near-infrared light to measure spiral amplitudes: for example, Iye *et al.* (1982); Considere & Athanassoula (1988); and Elmegreen *et al.* (1992). Although these and other studies made quantitative measurements of the strength, coherence, and multiplicity of the optical arms, the concern has persisted that the relationship between light and mass may be distorted by dust and by bright young stars formed in the arms. Recently, the advent of solid-state infrared detector arrays has allowed photometrically precise images at J, H, and K, where dust (probably) and young stars (arguably) are much less problematic than they are at visible wavelengths. In a detailed study of M51, Rix (1993) and Rix & Rieke (1993) have made a good case that $K$ light traces mass—or to be more cautious, that the extinction at $K$ is $\lesssim 10\%$ and that the nonaxisymmetric structure at $K$ appears not to be dominated by stars whose age is less than the interarm crossing time. These authors also found that the $I$-band surface brightness was very faithful to that at $K$. Quillen *et al.* (1994a) and Quillen *et al.* (1994b) have used JHK photometry to reconstruct the nonaxisymmetric potentials of the stellar bars in NGC 4314 and NGC 7479.

The object of the present study is to measure spiral-arm torques directly from photometry. This seems not to have been done before. The torques depend only upon the mass distribution, and not upon the persistence or pattern speed of the arms. One needs a photometric tracer of the mass. Although a dark-matter halo that is not at all well traced by the light is probably present and may dominate the rotation curve, this is not a serious concern for the torque measurement. Dynamical stability requires that the halo be much hotter kinematically and thicker geometrically than the visible stars [Ostriker & Peebles (1973)], so the halo probably does not participate in spiral structure.

To translate the torque into a secular evolution rate, one must assume that the torque measured today is typical of the galaxy at other times, although one need not assume that the spiral pattern is steady. Also, the torque has to be compared to the angular momentum of the stars in order to estimate how rapidly their orbits change, which requires knowledge of the rotation curve.

To illustrate the feasibility of the torque measurements, we have applied our methods to the bright and well-studied Virgo spiral M100. The data are described in §2, together with our assumptions about dust obscuration, mass-to-light ratios, and disk thickness. The method of analysis is laid out in §3, together with tests on artificial data. Results for M100 are given in §4, and §5 sums up.

## 2.  M100

### 2.1.  The Images

Images of M100 were taken from *A Catalog of Digital Images of 113 Nearby Galaxies* [Frei *et al.* (1995), henceforth FGGT]. Observations were made on the 1.5 meter telescope (P60) of the Palomar Observatory on the night of 4 May 1991 with the Wide Field Prime Focus Universal Extragalactic Instrument. Images were taken in the $g$, $r$, and $i$ bands of the Thuan-Gunn photometric system [Thuan & Gunn (1976), Wade *et al.* (1979)]. These filters are centered at 0.50 $\mu$, 0.65 $\mu$, and 0.82 $\mu$, respectively. See Frei & Gunn (1994) and FGGT for details.

The field of view is $16' \times 16'$, and it is projected on an $800 \times 800$ Texas Instruments CCD chip in the Cassegrain focus, yielding a scale of $1.''19$ pixel$^{-1}$. Using Hubble Space Telescope observations of cepheids, Freedman *et al.* (1994) have estimated the distance of M100 as $17.1 \pm 1.8$ Mpc. Thus the projected size of a pixel is $99 \pm 10$ pc.

A uniform sky level was subtracted from the images and foreground stars were removed [see Frei (1995)]. The photometric calibration by FGGT used observations of seven standard stars. We have converted the calibrated data in $gri$ to the corresponding Johnson bands $VRI$ according to the transformations of Frei & Gunn (1994). After transforming the colors and rectifying the images (see below), we have compared the azimuthally-averaged surface brightness profiles and total magnitudes with the photometry of de Jong & van der Kruit (1994). Thus, whereas we find $m_I = 8.38$, $m_R = 8.99$, and $m_V = 9.47$ (errors not estimated), de Jong & van der Kruit (1994) quote $m_I = 8.56 \pm 0.11$, $m_R = 9.01 \pm 0.10$, and $m_V = 9.56 \pm 0.08$. All magnitudes refer to the light within $R_{25}$, the radius of the 25th-magnitude isophote (at $B$). From the data of FGGT, we estimate $R_{25} = 221''$ or 21.9 kpc, in excellent agreement with the value $222''$ cited by the RC3 [de Vaucouleurs *et al.* (1991)]. Therefore, we believe that the photometry and color transformations available to us are sufficiently accurate for our purposes.

The images were rectified to correct for inclination. This is important, because tests on artificial data indicate that the torque estimate is sensitive to rectification errors (§3). There is no doubt that M100 is seen rather close to face-on, but published estimates of the position angle and inclination vary considerably. Having made the attempt ourselves, we are skeptical of estimates based on photometry alone. The ellipticity caused by inclination is second-order in the inclination ($i$) when $i$ is small; and it is entangled with the intrinsic $m = 2$ spiral structure. The observed velocities are first-order ($\propto \sin i$); and although spiral structure may disturb the velocity field, the perturbations are also proportional to $\sin i$. It is relatively easy to identify the kinematic axes. Therefore we have adopted the position angle 158° determined by Arsenault *et al.* (1988) from Hα Fabry-Perot data, which is in good agreement with other kinematic measurements [Rubin *et al.* (1980): 155°; Warmels (1988), Guhathakurta *et al.* (1988): 153°]. Some photometric determinations are almost orthogonal, however: de Vaucouleurs *et al.* (1991): 30°; Grosbøl (1985): $58 \pm 4°$.

We determine the inclination by requiring that the rotation curve agree with the Fisher-Tully (FT) relation. With the total magnitudes $m_R$ and $m_I$ cited above and the reddening corrections derived in §2.2, the relations of Pierce & Tully (1992) predict a circular velocity $V_c = 270 \pm 22$ km s$^{-1}$, where the error is the sum in quadrature of Freedman *et al.* (1994)'s uncertainty in the distance modulus, the 0.2 mag difference between our $m_I$ and that of de Jong & van der Kruit (1994), and the 0.1 mag intrinsic width of the FT relation in $R$ and $I$. Comparing this with the HI data of Guhathakurta *et al.* (1988), we find $i = 27° \pm 2.5°$.

We have used the central $512 \times 512$ pixels ($609'' \times 609''$) of each field for our calculations. Grayscale plots of the original $i$, $r$, $g$ images and the rectified $g$ image are given in Fig. 1. The $g$ image shows more small-scale structure than the $i$, presumably as a result of dust and recent star formation. The straight line in Fig. 1d shows the major axis (line of nodes) we have adopted. The small companion seen at the top of the image has proved to have negligible influence on the measured torque. Two prominent arms are clearly seen. They are more open in the rectified image than in the original.

## 2.2. Corrections for Galactic extinction and internal absorption

Since we want to use light as a tracer of mass, it is important to correct the data for the effects of dust.

According to Burstein & Heiles (1984), Galactic extinction of M100 in the B band is only $A_B^G \approx 0.04$mag. For a standard extinction curve [Savage & Mathis (1979)] that is compatible with Burstein & Heiles (1984)'s assumption $A_B/E(B-V) = 4$, we find $A_V^G \approx 0.031$, $A_R^G \approx 0.023$, $A_I^G \approx 0.014$ for the Galactic extinction at $V$, $R$, and $I$. These values are so small as to be essentially negligible compared to uncertainties in the stellar mass-to-light ratios, but we correct for them anyway.

The internal absorption within M100 can be estimated in several ways. Tully & Fouqué (1985) have derived an empirical formula for internal absorption, which is based on the trend in $B - H$ color with inclination in a sample of late-type spirals. Their formula predicts $< A_B^i > \approx 0.30$ mag for $i = 27°$ (and 0.27 mag for $i = 0°$). The reduction in surface brightness due to the dust, which is what these numbers measure, is not directly proportional to the extinction optical depth of the disk. The relationship between surface brightness and extinction depends on the relative thicknesses of the star and gas layers, and on scattering. Ignoring these fine points (because the final correction is small), we use the standard extinction law to derive $(A_V^i, A_R^i, A_I^i) = (0.23, 0.18, 0.11)$ mag. We can check this against Freedman *et al.* (1994)'s value for the mean extinction of 20 Cepheids in M100, as derived by comparing the period-luminosity relations in $V$ and $I$: these authors have obtained $\langle A_V \rangle = 0.15 \pm 0.17$. Finally, Xu & Buat (1994) have used the far-infrared emission of 135 spirals to deduce mean and median total face-on optical depths at $B$ of $0.60 \pm 0.04$ and 0.49, respectively. The surface-brightness reduction corresponding to these optical depths depends on details of the radiative transfer but is probably about half as large; thus, $\langle A_B \rangle \approx 0.3$ mag. We regard these three ways of estimating the internal absorption as roughly consistent.

Our adopted values for the total corrections to the surface brightness, accounting for both Galactic extinction and internal absorption, are $A_V = 0.26$ mag, $A_R = 0.20$ mag, and $A_I = 0.12$ mag. Hence the redenning-corrected total magnitudes of M100 are estimated to be $m_I = 8.26$, $m_R = 8.79$, $m_V = 9.21$.

## 2.3. Stellar Mass-to-Light Ratios

The mass-to-light ratio of the stellar population in M100 (or any other spiral) is the greatest uncertainty in estimating spiral-arm torques from photometry alone. We have used published population-synthesis models for this purpose.

Following the usual convention, we define $(M/L)_X$, the mass-to-light ratio in band $X$, as the mass of the stellar population in units of $M_\odot$ divided by the number of solar-type stars that would produce the observed light in band $X$, after correcting for absorption. Thus if $M$ is the mass of the population, $\mathrm{M}_X$ its absolute magnitude in band $X$, and $\mathrm{M}_{X,\odot}$ the corresponding absolute magnitude of the Sun, then

$$\frac{M}{M_\odot} = \left(\frac{M}{L}\right)_X 10^{0.4(\mathrm{M}_{X,\odot} - \mathrm{M}_X)} \tag{1}$$

The models of Arimoto & Jablonka (1991) as interpreted by Jablonka & Arimoto (1992) predict $(M/L)_H = 2.15$ for Sc galaxies, including the mass in associated gas and stellar remnants, but not halo dark matter. This requires assumptions about the slope and extent of the initial mass function and about the history of the star-formation rate. The mass-to-light ratios at $V, R, I$ can be derived from $(M/L)_H$ using the predicted colors $(V - H)_{\mathrm{Sc}} = 2.196$, $(R - H)_{\mathrm{Sc}} = 1.567$, and $(I - H)_{\mathrm{Sc}} = 0.989$, and the colors of the Sun [Worthey (1994)], $(V - H)_\odot = 1.47$, $(R - H)_\odot = 1.11$, and $(I - H)_\odot = 0.76$. Thus: $(M/L)_V = 4.20$, $(M/L)_R = 3.28$, and $(M/L)_I = 2.65$.

For comparison, the single-burst models of Worthey (1994) predict $(M/L)_I$ monotonically increasing from 1 to 3 with ages from 1.5 to 10 Gyr. According to these models, the dependence on metallicity is very weak in the $I$ band. For populations of age 5 Gyr and solar metallicity, $(M/L)_V = 3.1$ and $(M/L)_R = 2.6$. These bands are more sensitive to age and considerably more sensitive to metallicity than the $I$ band. All of Worthey's mass-to-light ratios are based on the initial mass in stars less massive than $2M_\odot$. We refer the reader to the papers of Arimoto & Jablonka (1991) and Worthey (1994) for further details.

Both the absorptions cited in the last subsection and the mass-to-light ratios cited in this one are mean values for the entire disk. The absorption is larger than the mean in the arms, however, where the gas and dust concentrate. Furthermore, the arms contain a higher proportion of young, luminous stars. Both of these effects, which have opposing influences on the effective mass-to-light ratio, are expected to be more severe at $V$ and $R$ than at $I$. When calculating the surface mass distribution from the photometry, we have *not* corrected for these variations in mass-to-light ratio. The extent to which our conclusions are compromised by mass-to-light variations can be gauged by comparing our results for the mass distribution and the torque estimated separately from the data in the three bands.

## 2.4. Thickness of the Disk

Infrared photometry of edge-on spirals indicates that the light profile perpendicular to the disk is constant with radius and well fit by an exponential law,

$$p(z) = \frac{\exp(-|z|/h_z)}{2h_z}, \tag{2}$$

with a vertical scale height $h_z$ that is of order one tenth the horizontal disk scale length, $h_r$.

In their infrared study of IC2531, Wainscoat *et al.* (1989) found $h_z \approx h_r/12$. In a similar study of N5907, Barnaby & Thronson (1992) fit a $\mathrm{sech}(z/h_z)$ profile—suggested by van der Kruit (1988) as a compromise between the exponential law (2) and the dynamically simpler $\mathrm{sech}^2(z/h_z)$ profile—and found $h_z/h_r \approx 1/9.3$. Kuijken & Gilmore (1989) fit their counts of $K$ dwarfs in the solar neighborhood to a double exponential: 82% of the stars are in a component with $h_z = 249$pc, and the rest have $h_z' = 1.00$kpc; they assume $h_r = 4.5$kpc, so $h_z/h_r \approx 1/18$ and $h_z'/h_r \approx 2/9$.

For the purpose of computing the torque, we have assumed the profile (2) with $h_z = h_r/12$.

## 3.  Recipe for the Torque

In this section we explain our methods for calculating the torque from the photometric data. We assume for now that we know the mass-to-light ratio and can convert the surface brightness, $\sigma$, to the stellar mass per unit area, $\Sigma$.

Let $(r, \theta, z)$ be cylindrical coordinates such that $r$ is the distance from the rotation axis of the galaxy, $z$ is the height above the equatorial plane, and $\theta$ is an azimuthal angle. The $z$ component of the gravitational torque across radius $R$, exerted on the outer $(r > R)$ part of the galaxy by the inner $(r < R)$ part, is

$$\begin{aligned}
\Gamma(R) &= G \int\limits_{r_1 > R} d^3\boldsymbol{x}_1 \rho(\boldsymbol{x}_1) \int\limits_{r_2 < R} d^3\boldsymbol{x}_2 \rho(\boldsymbol{x}_2) \frac{(\boldsymbol{x}_1 \times \boldsymbol{x}_2)_z}{|\boldsymbol{x}_1 - \boldsymbol{x}_2|^3} \\
&= G \int\limits_{-\infty}^{\infty} d\varepsilon \, C(\varepsilon) \int\limits_{r_1 > R} d^2\boldsymbol{r}_1 \Sigma(\boldsymbol{r}_1) \int\limits_{r_2 < R} d^2\boldsymbol{r}_2 \Sigma(\boldsymbol{r}_2) \frac{(\boldsymbol{r}_1 \times \boldsymbol{r}_2)_z}{[|\boldsymbol{r}_1 - \boldsymbol{r}_2|^2 + \varepsilon^2]^{3/2}}, \tag{3}
\end{aligned}$$

where $\boldsymbol{x}_i$ is the vector corresponding to coordinates $(r, \theta, z)$, $\boldsymbol{r}$ is the projection of this vector onto the equatorial plane, $\rho(\boldsymbol{x})$ is the stellar mass per unit volume, and $\Sigma(\boldsymbol{r})$ is the mass per unit area. In the second line, we have assumed the disk has a constant vertical structure, so that $\rho(\boldsymbol{r}, z) = \Sigma(\boldsymbol{r})p(z)$ with $\int dz \, p(z) = 1$, and we have introduced the

vertical autocorrelation

$$C(\varepsilon) \equiv \int\limits_{-\infty}^{\infty} dz \ p(z)p(z+\varepsilon), \quad \int\limits_{-\infty}^{\infty} d\varepsilon \ C(\varepsilon) = 1. \tag{4}$$

It is convenient to replace the integration over $\varepsilon$ in Eq. (3) by evaluation at $\varepsilon = \varepsilon_o$, for an appropriate choice of $\varepsilon_o$. Taking $\varepsilon_o = 0$ is equivalent to treating the disk as though it had zero thickness. In fact the disk has some characteristic thickness $h_z \sim 2/p(0)$. In the limit $kR \gg 1$, $\Gamma(R)$ can be decomposed into independent contributions from each Fourier component $\Sigma_{\boldsymbol{k}}(\boldsymbol{r}) \propto \exp(i\boldsymbol{k} \cdot \boldsymbol{r})$. For fixed amplitude, the contribution from any such component is a decreasing function of the thickness: if $kh_z \ll 1$, then the thickness hardly matters, but if $kh_z \gtrsim 1$, the reduction in the torque is substantial. Thus if $p(z)$ has the exponential form (2) then it can be shown that the reduction factor is $(1 + |kh_z|)^{-1}$. If one replaces $C(\varepsilon) \to \delta(\varepsilon - \varepsilon_o)$, however, then the reduction factor becomes $\exp(-|k\varepsilon_o|)$. Our prescription for $\varepsilon_o$ is that the exponential function should match the true reduction factor when the latter falls to $1/2$. In the case of the profile (2), this leads to $\varepsilon_o = h_z \ln(2) \approx 0.693 h_z$.[1]

We decompose the surface mass distribution into its Fourier components over azimuthal angle $\theta$ at each radius

$$\Sigma(r, \theta) = \Sigma_0(r) + \sum_{m=1}^{\infty} \Sigma_{cm} \ \cos(m\theta) + \Sigma_{sm} \ \sin(m\theta). \tag{5}$$

The softened newtonian potential can be expanded in a series of modified Laplace coefficients:

$$\frac{1}{\sqrt{r_1^2 + r_2^2 - 2r_1 r_2 \cos(\theta_1 - \theta_2) + \varepsilon_o^2}} = \sum_{m=0}^{\infty} F_m(r_1, r_2; \varepsilon_o) \cos m(\theta_1 - \theta_2), \tag{6}$$

in which [Prasad (1932)]

$$F_m(r_1, r_2; \varepsilon_o) = \frac{2 - \delta_{m0}}{\pi \sqrt{r_1 r_2}} \ Q_{m-1/2} \left( \frac{r_1^2 + r_2^2 + \varepsilon_o^2}{2 r_1 r_2} \right). \tag{7}$$

Here $Q_{m-1/2}(x)$ is a Legendre toroidal function of the second kind and $\delta_{ij}$ is the Kronecker delta.

---

[1] The isothermal profile $p(z) = \mathrm{sech}^2(z/h_z)/2h_z$ is sometimes used. In this case the reduction factor is $[1 - \zeta + 1/2\zeta^2\psi(2, 1 + 1/2\zeta)]$, where $\zeta \equiv kh_z$ and $\psi(n, z)$ is the polygamma function, and $\varepsilon_o \approx 0.7812 h_z$.

We can now integrate Eq. (3) over azimuthal angles to obtain

$$\Gamma(R) = \sum_{m=1}^{\infty} \Gamma_m(R), \tag{8}$$

$$\Gamma_m(R) = G\pi^2 m \int_R^{\infty} r_1 \, dr_1 \int_0^R r_2 \, dr_2 \, F_m(r_1, r_2; \varepsilon_o) \left[ \Sigma_{sm}(r_1)\Sigma_{cm}(r_2) - \Sigma_{sm}(r_2)\Sigma_{cm}(r_1) \right].$$

The torque vanishes if the surface density is bilaterally symmetric around any horizontal axis, since if we were then to orient our coordinates so that the symmetry axis lay along $\theta = 0$, then $\Sigma_{sm}(r) \equiv 0$ for all $m$ and $r$. Whatever the shape of the galaxy, there is no contribution to the torque from any $m = 0$ component of the density, such as an axisymmetric dark halo.

The torque due to a spiral arm of the form

$$\Sigma_m(r, \theta) = \Sigma_{m0}(r) \cos(m\theta + \phi_m(r)) \tag{9}$$

approaches asymptotically

$$T_m(R) \approx \frac{G\pi^2 m \beta}{(\beta^2 + m^2)^{3/2}} \, e^{-k\varepsilon_o} \, R^3 \Sigma_{m0}^2(R), \qquad \beta \equiv \left. \frac{d\phi_m}{d\ln r} \right|_{r=R}, \tag{10}$$

in the limit

$$kR \equiv \sqrt{\beta^2 + m^2} \gg \max\left(1, \frac{d\ln\Sigma_{m0}}{d\ln r}, \frac{d\ln\beta}{d\ln r}\right). \tag{11}$$

Constant $\beta$ gives a logarithmic spiral of pitch angle $\tan^{-1}(m/\beta)$. At a given $k$, the torque is maximized when $\beta = m$, which corresponds to a very loosely-wrapped spiral with pitch angle = $45°$.

The contribution of the disk to the rotation curve can be expressed in terms of the $m = 0$ component of the surface density:

$$V_{c,disk}^2(r) = -2\pi G r \frac{\partial}{\partial r} \int_0^{\infty} \Sigma_0(r') F_0(r, r'; \varepsilon_o) r' dr'. \tag{12}$$

The actual rotation curve, $V_c$, includes contributions from the bulge, spheroid, and dark halo, and corrections for asymmetric drift.

Given $V_c(r)$, we estimate the angular momentum of the disk interior to radius $R$:

$$J(R) = 2\pi \int_0^R r \, V_c(r) \, \Sigma_0(r) \, r \, dr. \tag{13}$$

The angular momentum within $R$ changes because of the torque across $R$ and because of changes in the mass within $R$:

$$\frac{\partial J}{\partial t}(R,t) = -\Gamma(R,t) + \frac{\partial J}{\partial M}\frac{\partial M}{\partial t}(R,t). \qquad (14)$$

Hence, a characteristic evolution time at radius $R$ is

$$t_{\rm ev}(R) = \left|\frac{J(R,t)}{\Gamma(R,t)}\right|. \qquad (15)$$

We have tested the above procedures against artificial data for $m = 2$ logarithmic spirals with two radial profiles: a power-law disk $[\Sigma_2(r) \propto r^{-1}]$ and an exponential disk $[\Sigma_2(r) \propto \exp(-r/h_r)]$. Like the real images, our artificial data is binned into $512^2$ pixels. The asymptotic expression Eq. (10) agrees very well with the integral formula Eq. (8) for the exponential density profile; the relative error is a few percent when the "image" subsumes about four scale lengths (as do the real data—see §4). But the power-law disk is very sensitive to truncation of the image. Whereas the correct value of $\Gamma(R)$ should increase linearly with radius, assuming unlimited radial extent and zero thickness, the torque calculated from the artificial data oscillates with increasing amplitude around the correct value.

Substantial errors occur in the torque calculated from Eq. (8) if the image is incorrectly centered or rectified. An incorrect center gives rise to a very strong $m = 1$ harmonic, and the torque oscillates strongly with radius around its correct value. Improper inclination amplifies the torque vastly and also produces some variations with radius. This does not seem to be a severe problem in the case of M100, as we discuss in §4.3.

## 4. Results

### 4.1. Photometric results

We have applied the techniques described in §3 to M100. We decompose the surface light distribution into its Fourier harmonics, $\sigma_{cm}(r)$ and $\sigma_{sm}(r)$, for $m \leq 10$. Since the Fourier components (see Fig. 4 below) decrease rapidly with $m$, and their contributions to the torque decrease even faster [cf. Eq. (10)], the first ten harmonics are more than sufficient to reproduce the relevant features of the surface density. Fig. 2 shows images reconstructed from these Fourier components in $i$ and $g$. The reconstructions are faithful to the originals, but smoother.

The axisymmetric surface-brightness profile, $\sigma_0(r)$, is shown in Fig. 3 in semilogarithmic form. In this and all subsequent figures, we have transformed the photometric data into the Johnson $VRI$ bands, as described in §2.1. A compact component, presumably the bulge dominates the light within $20''$. Between $60''$ and $160''$, the disk can be approximated by an exponential with scale length $h_r \approx 64''$ (5.35 kpc). The three bands are parallel in this diagram to better than 0.3 mag between $10''$ and $190''$; at larger radii $I$ declines more rapidly than $V$ or $R$.

Profiles of the nonaxisymmetric Fourier components are shown in Fig. 4. The moduli and phases are very consistent across the three photometric bands within $200''$, especially for $m \geq 2$.

Figure 5 shows the azimuthal variation of surface brightness at five selected radii. Following Rix (1993, Fig. 8), we have subtracted all of the odd Fourier harmonics. While it is clear that these profiles are not just $m = 2$ sinusoids, the differences among the three bands are rather small. The maximum brightness contrast between arm and interarm regions is about 1.2 mag, or a factor of 3, in the outer parts of M100; thus we are in qualitative agreement with Elmegreen *et al.* (1989), who found contrasts as large as a factor of 4.

## 4.2. The rotation curve

We have used Eq. (12) to calculate the contribution of the disk to the rotation curve; the results for all three photometric bands are shown in Fig. 6. A constant vertical scale height $h_z = 446$ pc has been assumed even in the "bulge" (cf. §2.4). Because of the close similarity of the axisymmetric light profiles (Fig. 3), the three rotation curves are similar in shape, but with Jablonka & Arimoto (1992)'s mass-to-light ratios, they differ in amplitude. Therefore, we have fixed $(M/L)$ in each band so as to minimize the differences among the three curves in the range $60'' < R < 200''$ region) and to cause them to peak at about 270 km s$^{-1}$ (§2). The required ratios of *dynamical* mass to light are 2.85, 3.49, and 3.80 in $I$, $R$, and $V$, respectively (in solar units, after correction for dust).

The dynamical mass includes the mass in the dark halo within the radius where the rotation curve peaks ($R_{max} \approx 160''$). Recall that the corresponding stellar $(M/L)$'s predicted by the population-synthesis models are 2.65, 3.28, and 4.20 (§2.3). Thus the disk would appear to account for a fraction $f \approx 0.93$ (based on $I$) of the mass contributing to the observed part of the rotation curve. Although the predicted colors are in fair agreement with the observations—e.g. $(V - I)_{Sc} = 1.21$ versus $(V - I)_{M100} = 0.95$ (or 1.09 before the reddening correction)—-the models of Arimoto & Jablonka (1991) and Worthey (1994) may overestimate the mass in stellar remnants and very-low-mass stars, which would contribute little light in any of our bands. Rotation-curve fitting [Persic & Salucci (1990)] indicates $f \approx 0.5$, a value typical of bright spirals. The bar-instability criterion of Ostriker & Peebles (1973), if applicable, also requires $f \lesssim 0.5$.

## 4.3. The Torque and the Evolution Time

The torque computed from all three bands is shown in Fig. 7. Light has been translated to mass according to the dynamical ratios above, reduced by a factor $f = 0.93$. Because it is quadratic in the surface density, the torque scales as $f^2$. The agreement among the three bands is clearly very good. As shown by the lower panel, almost all of the torque is created by the $m = 2$ Fourier components. The torque peaks at $R_{\rm peak} \approx 130''$, or about 2 scale lengths. It falls off at larger radii because the disk as a whole quickly fades, although the arm-interarm contrast remains strong. The local rate of accumulation of angular momentum is proportional to $-d\Gamma/dR$, so the disk interior to $R_{\rm peak}$ is (at present) losing angular momentum, while that exterior is gaining it. The slightly negative values of $\Gamma$ at $R \approx 80''$ may be due to errors in centering or rectifying the image. It is also possible that we are picking up a weak leading wave (assuming that the spiral structure is predominantly trailing), whose presence would be required in steady-state by the swing-amplifier mechanism [e.g. Toomre (1981)].

As we indicated in §3, the measured torque curves are sensitive to the rectification of the galaxy. The upper panel of Fig. 8 shows the results of tests on an artificial galaxy with an exponential profile and an $m = 2$ logarithmic spiral. The hatched region shows the range swept out by the "measured" torque when we varied the assumed position angle by $\pm 10°$ and the inclination by $\pm 5°$ around the true values. The data for M100 are less sensitive to the rectification parameters. In the lower panel of Fig. 8 the $I$-band torque is plotted for our preferred values, $i = 27°$ and $PA = 158°$. The hatched region is the range swept out when the $PA$ is varied by $\pm 10°$ and $i$ by $\pm 5°$. Our tests on artificial data show that the sensitivity to rectification increases as the spiral arms become more tightly wrapped. M100 is a relatively loose spiral, hence relatively insensitive. The artificial spiral used in constructing the upper panel was chosen to have a pitch angle comparable to that of M100 ($\beta \approx 1.7$).

Figure 9 depicts the evolution time Eq. (15) for the three bands. Because of the way $t_{\rm ev}(R)$ has been defined, it must diverge at large radii, where the torque vanishes, and it isn't reliable at small radii, where the spiral structure is poorly defined. Therefore, the most interesting features of these curves are their minima, which occur at about two disk scale lengths: $t_{ev,min} \approx 6$ Gyr. In terms of the mass fraction $f$ introduced above, $J(R) \propto f$ for a fixed amplitude of the rotation curve, so $t_{ev} \propto f^{-1}$. Thus if $f \approx 0.5$ instead of 0.93, the values shown in Fig. 9 should be increased by a factor 1.9.

## 5. Discussion and conclusions

We have shown that a gravitational angular-momentum flux, or torque, can be measured directly from the mass distribution in spiral disks. Using multiband photometry, we have carried out this measurement for M100. The estimated torque implies a dynamical evolution time $5 - 10 \times 10^9$ yr, with the main source of error being the stellar mass-to-light ratio. Our measurement is somewhat sensitive to the orientation of the galaxy, which we have estimated from external kinematic data. Dust and young stars appear not to be problematic. We obtain consistent results in all three photometric bands. Although dust lanes and local enhancements in the surface brightness due to young stars are more prominent in the shorter-wavelength images, these features do not affect the $m = 2$ Fourier components or the torque very much, presumably because they cover rather small areas.

M100 may not be a typical spiral. It is bright, and its grand-design structure may have been stimulated by gravitational perturbations from its small companions. The "bulge" is not well fit by a de Vaucouleurs profile, and it may harbor a small bar [Shaw *et al.* (1995)]. Therefore, the implications discussed in the next paragraph need to be confirmed by applying these methods to a fair sample of spirals.

We cannot resist the temptation to speculate. If the spiral structure we see today is typical of M100's past, it has very substantially altered the radial distribution of the oldest stars in this galaxy. Whereas the self-similar infall model of Gunn (1981) predicts that the disk scale length should increase with time as a consequence of the arrival of higher-angular-momentum gas, the torques discussed here would decrease the scale length by removing angular momentum from most of the pre-existing stars. It is unknown which of these two effects dominates, but the implications for redshift-angular-diameter tests based on spirals are not encouraging. More positively, we can say that our results are consistent with the speculation of Toomre (1981) that spirals stabilize themselves against nonaxisymmetric modes by redistributing mass. Perhaps spirals are enormous accretion disks, and because they are so well-resolved, they may offer lessons that can be applied to smaller self-gravitating disks such as those believed to exist around protostars.

We would like to thank James Gunn and James Rhoads for discussions. We are indebted also to Edward Fitzpatrick for help with the IRAF image processing software. This work was supported by NASA's Astrophysics Theory Program under grant NAGW-2419.

Fig. 1.— (a), (b), (c): Images of M100 in $i$, $r$, and $g$ bands, respectively. Scale bar on right = 1′. (d) Rectified $g$ image, using $i = 27°$, $PA = 158°$. Straight line across image drawn at 158°. In all panels north is up and east is left.

Fig. 2.— Comparison of original data to those reconstructed from the first 10 Fourier components. (a) Rectified $i$ image, (b) Fourier reconstruction for $i$. (c), (d): Like (a) and (b), but for $g$. Scale as in Fig. 1, but note different orientation.

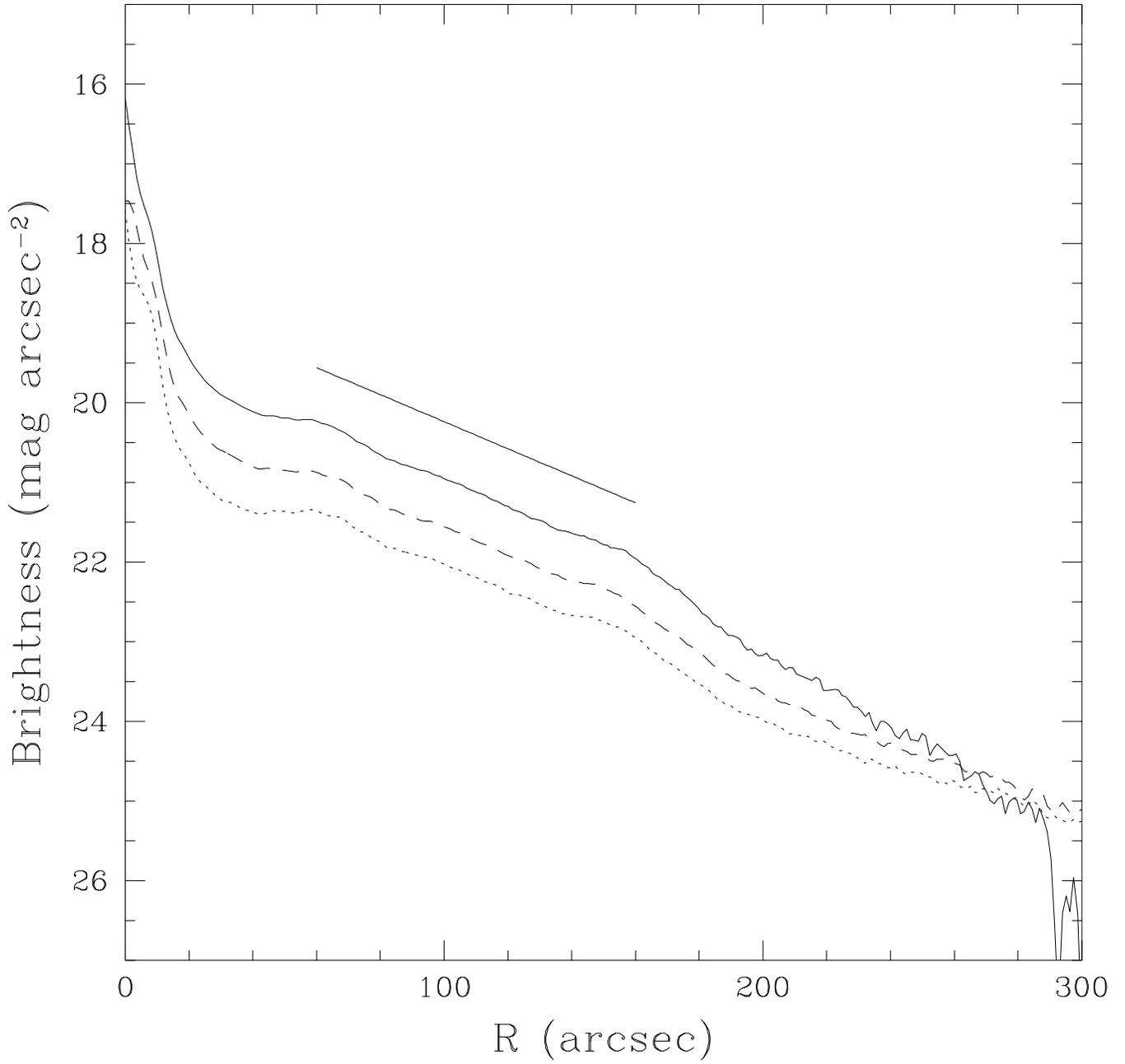

Fig. 3.— Axisymmetric surface brightness profiles converted to Johnson photometric bands. Solid: $I$. Dashed: $R$. Dotted: $V$. Straight line is an exponential with scale length 64″.

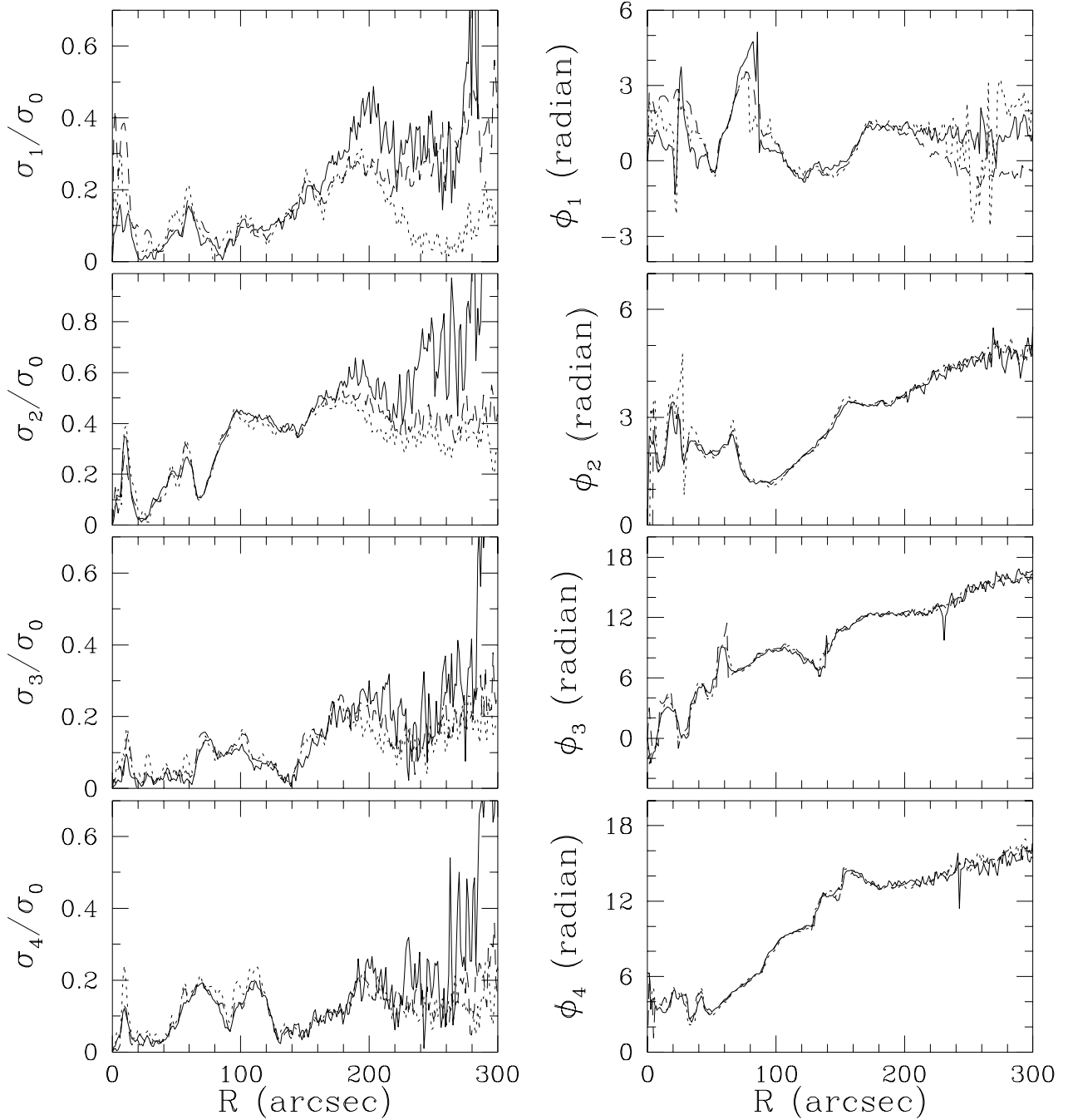

Fig. 4.— Fourier components $m = 1, 2, 3, 4$ of surface brightness. Left panels are moduli normalized to axisymmetric component: $(\sigma_{cm}^2 + \sigma_{sm}^2)^{1/2} / \sigma_0$. Right panels are the phases, $\tan^{-1}(-\sigma_{sm}/\sigma_{cm})$. Solid lines: $I$; dashed: $R$; dotted: $V$.

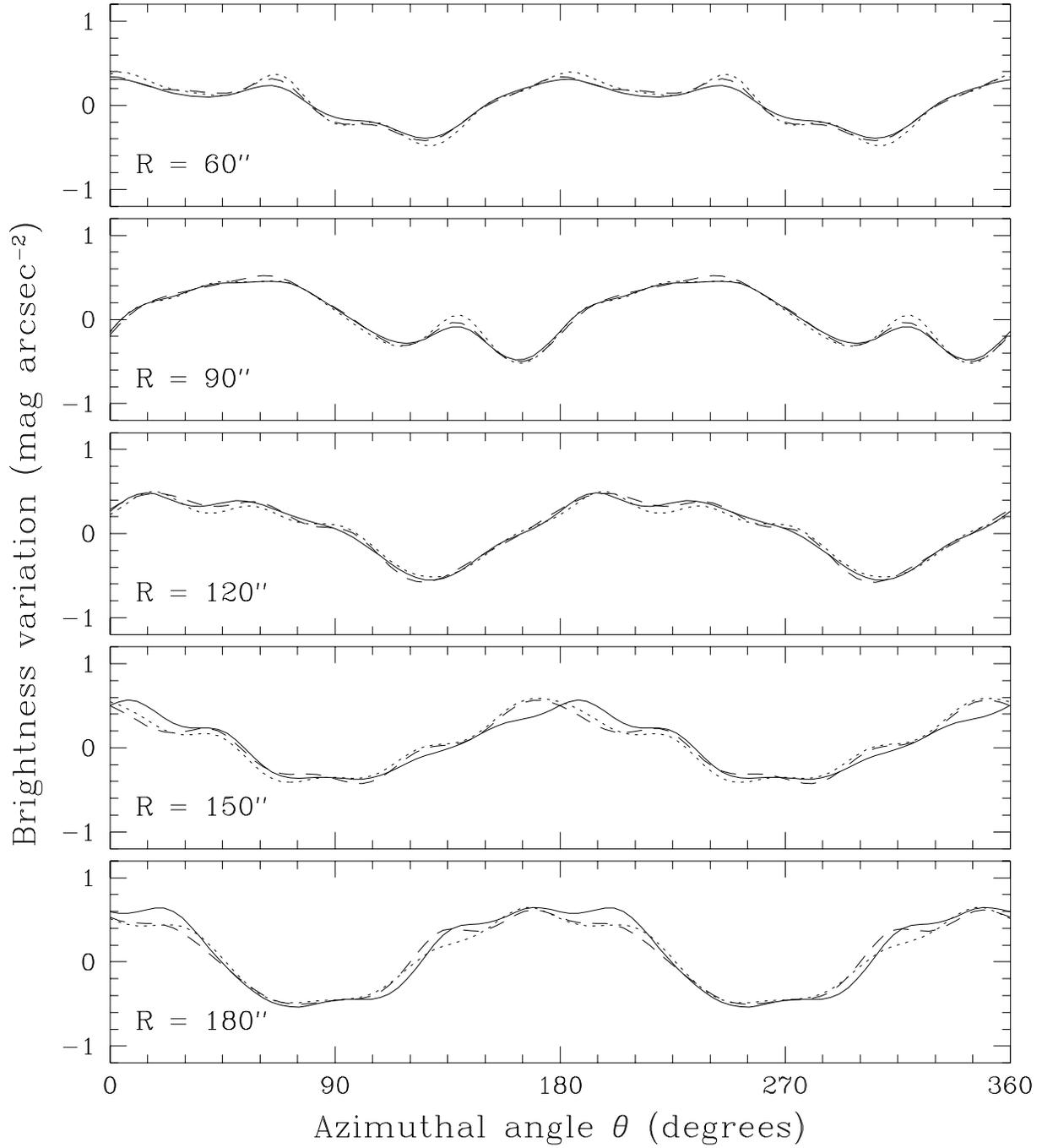

Fig. 5.— Arm - interarm density contrast. Brightness in magnitudes relative to azimuthal mean, plotted against azimuth at 5 radii, after subtraction of odd Fourier harmonics. Solid lines: $I$; dashed: $R$; dotted: $V$.

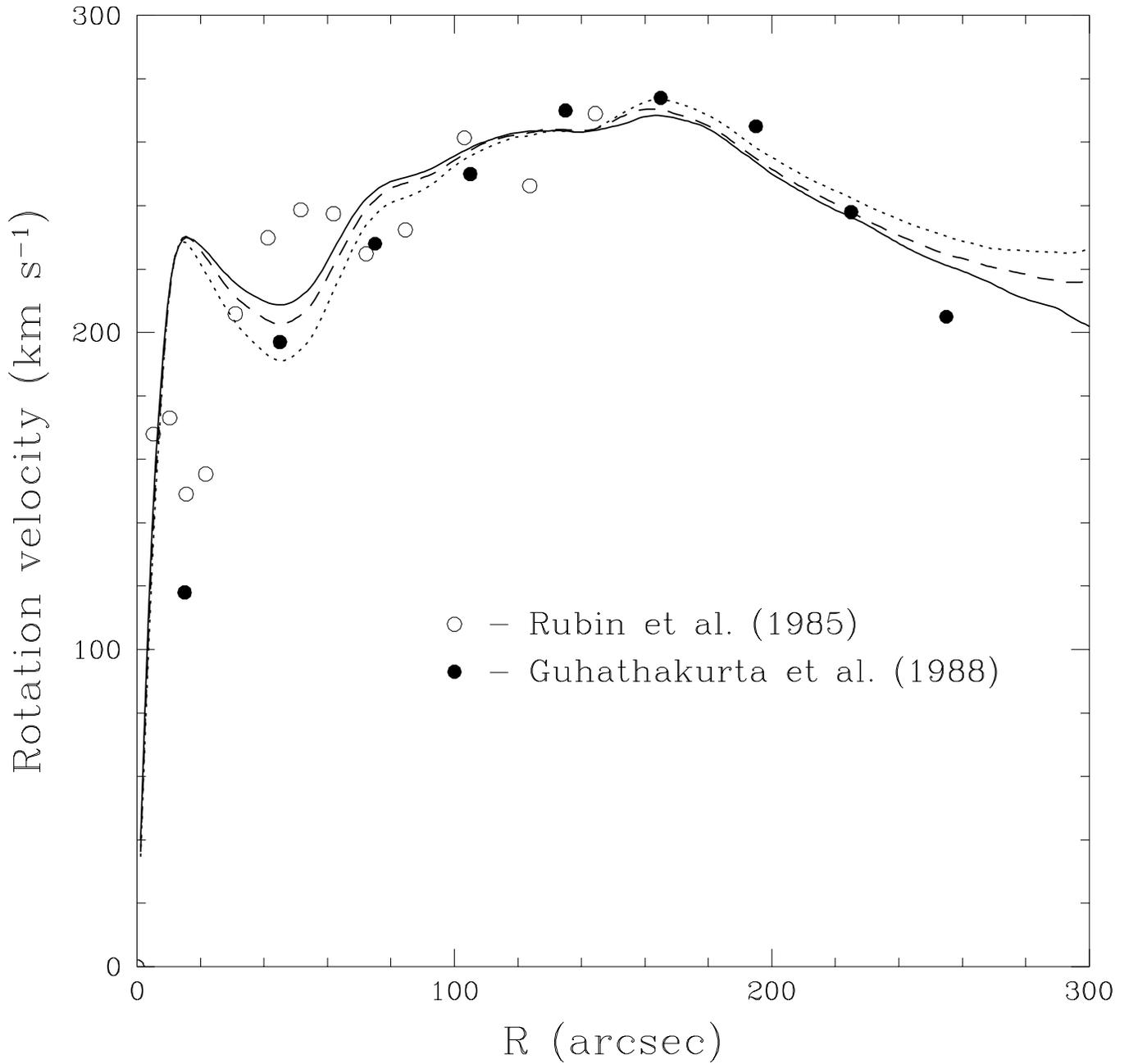

Fig. 6.— Rotation curves for M100 calculated from photometry. Solid line: $I$; dashed: $R$; dotted: $V$. Symbols: Velocity data from the the indicated sources; Rubin *et al.* data rescaled from $i = 35°$ to $i = 27°$.

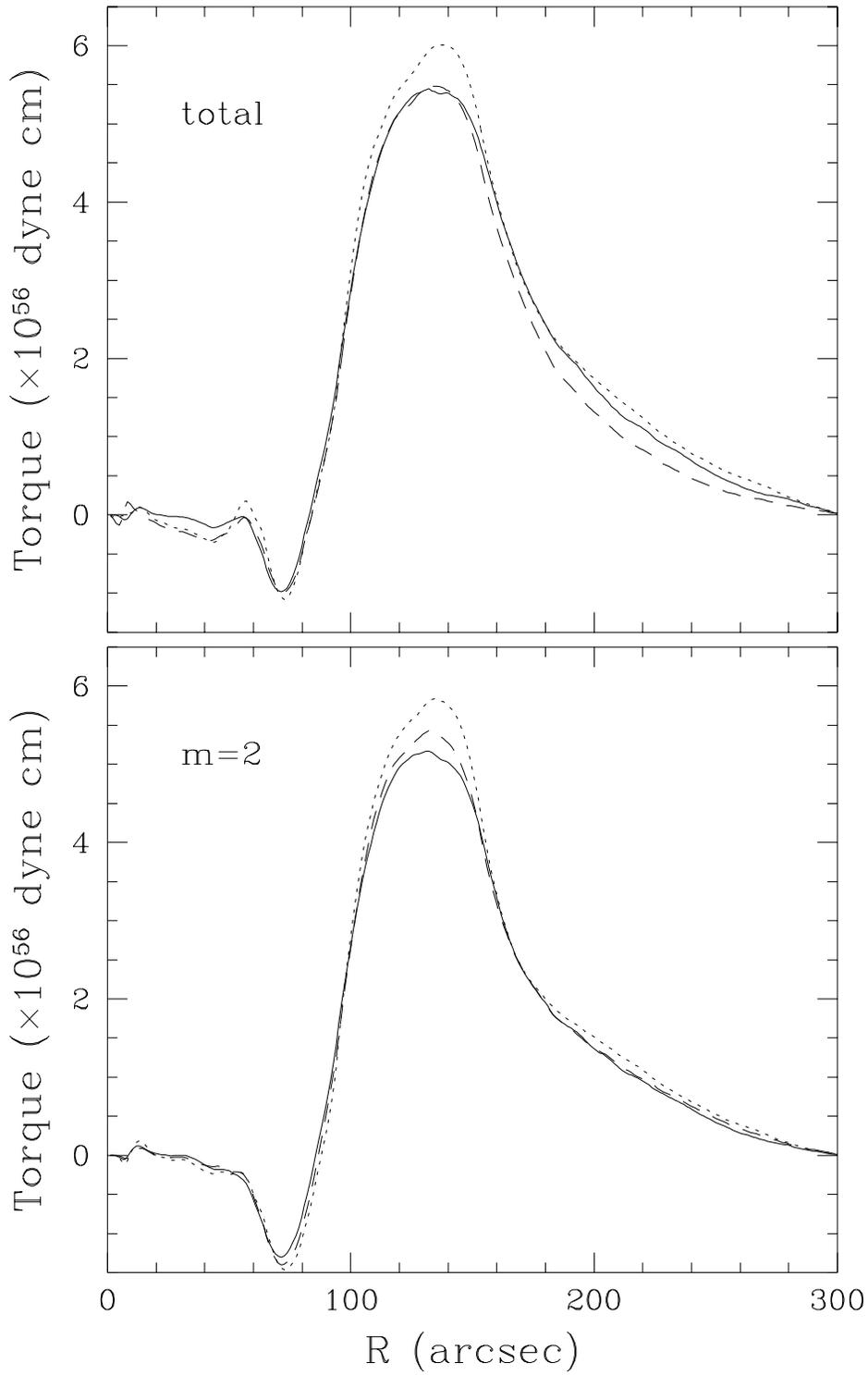

Fig. 7.— Torque versus radius for all three bands. Upper panel includes contributions from $1 \leq m \leq 10$, lower from $m = 2$ only. Solid lines: $I$; dashed: $R$; dotted: $V$.

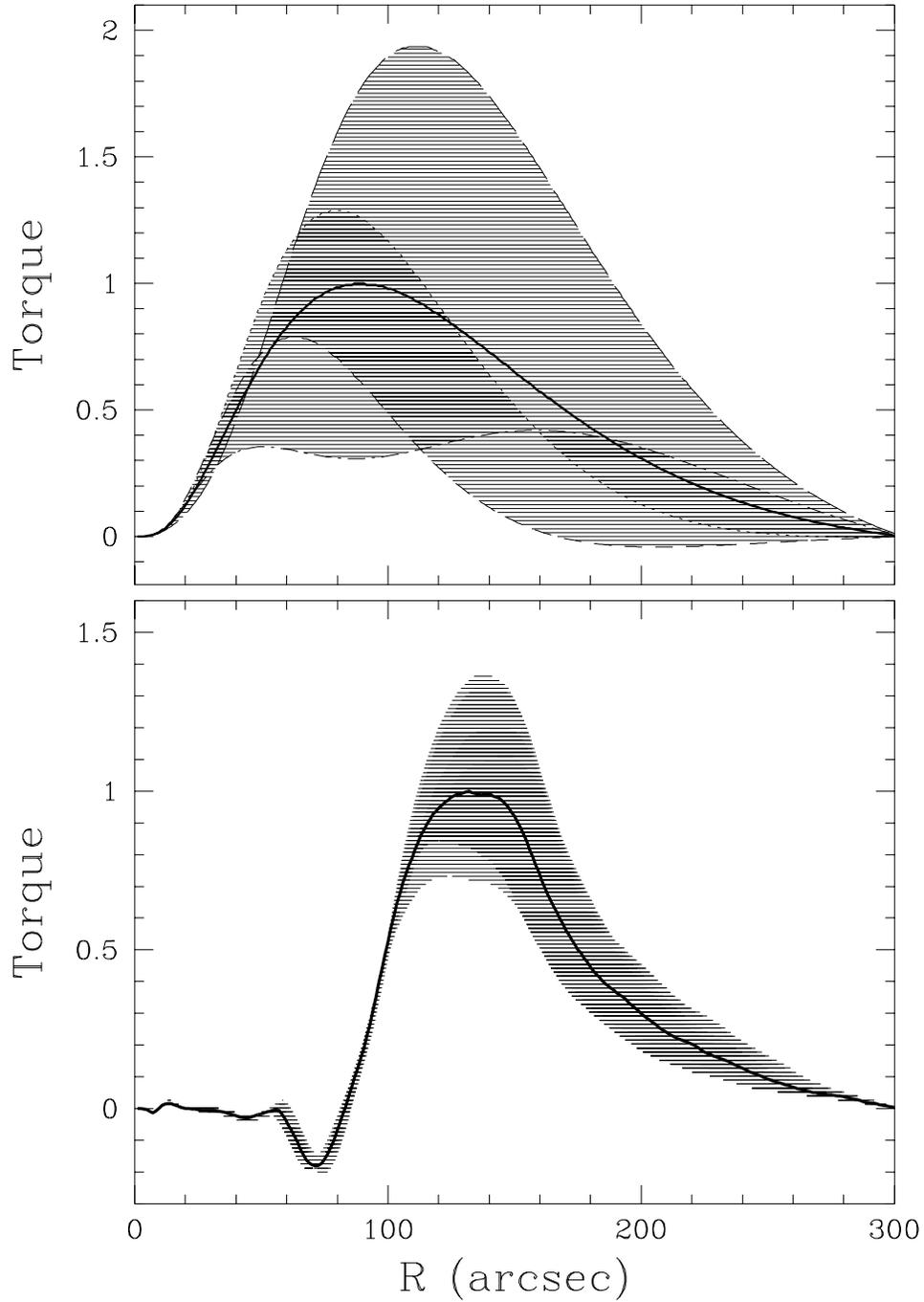

Fig. 8.— Sensitivity of the torque to rectification parameters. All curves rescaled by the same factor such that correct (solid) curve has peak value unity. Upper panel: artificial galaxy. Lower panel: M100. For full explanation, see text.

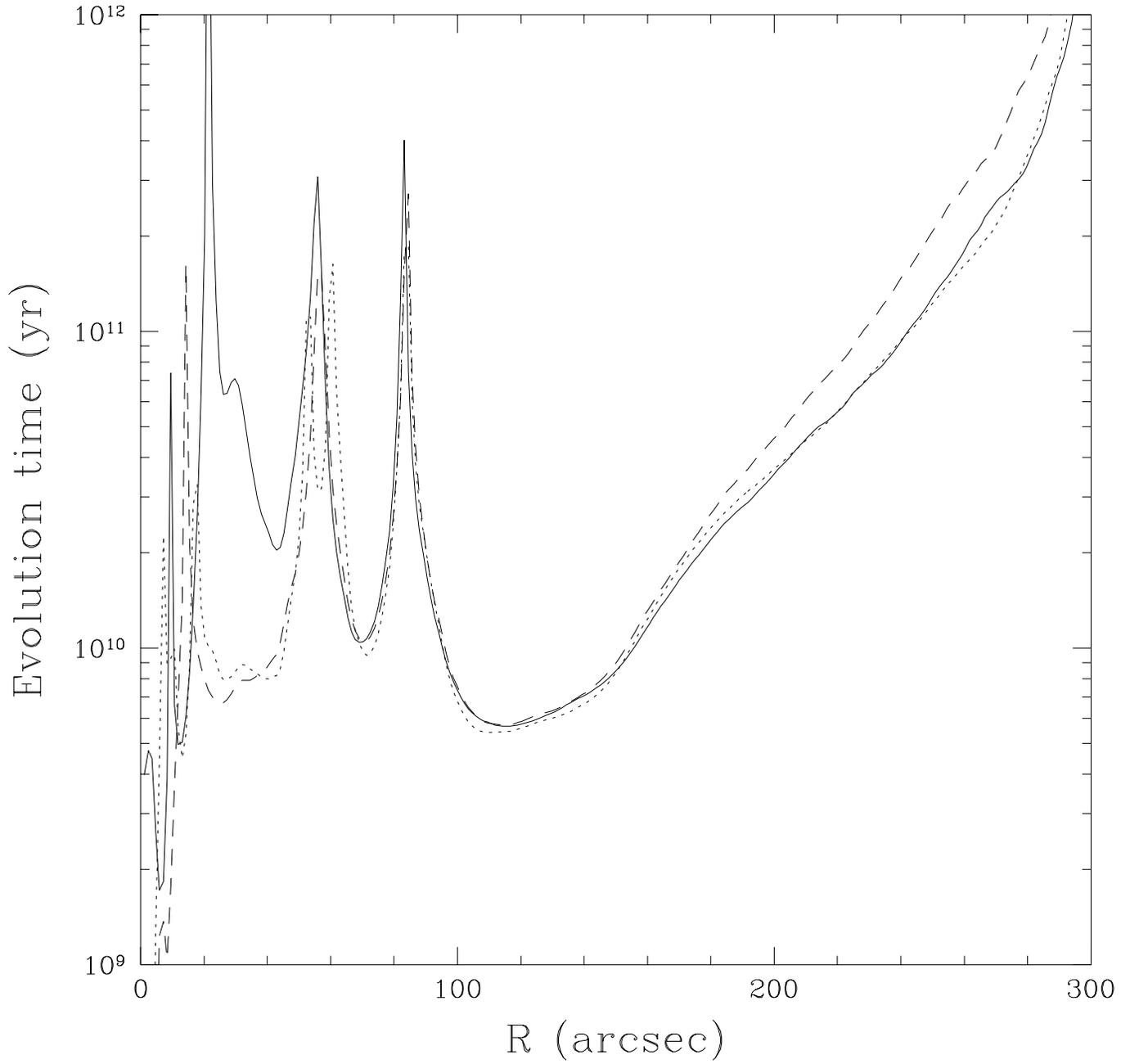

Fig. 9.— Evolution time Eq. (15). Solid lines are for $I$, dashes are for $R$, and dots are for $V$.